\documentclass[aps,twocolumn,showpacs]{revtex4}
\usepackage{amsfonts,amsmath,amssymb,latexsym,epsfig}
\begin{document}
\title{Transmission and Reflection in the Stadium Billiard:\\ Time-dependent asymmetric transport}
\author{Carl P. Dettmann \footnote{Carl.Dettmann@bristol.ac.uk}}
\affiliation{School of Mathematics, University of Bristol, United Kingdom}
\author{Orestis Georgiou \footnote{maxog@bristol.ac.uk}}
\affiliation{School of Mathematics, University of Bristol, United Kingdom}

\begin{abstract}
We investigate the transmission and reflection survival probabilities for the chaotic stadium billiard with two holes placed asymmetrically. Classically, these distributions are shown to have algebraic or exponential decays depending on the choice of injecting hole and exact expressions are given for the first time and confirmed numerically. As there is no reported quantum theoretical or experimental analogue we propose a model for experimental observation of the asymmetric transport using semiconductor nano-structures and comment on the relevant quantum time-scales.
\end{abstract}
\maketitle
Billiards \cite{Tabachnikov} are systems in which a particle alternates between motion in a straight line and specular reflections from the walls of its container, while open billiards contain one or more holes through which particles may escape.
Billiards demonstrate a broad variety of behaviours including regular, chaotic and mixed phase space dynamics, depending on the geometry, whilst allowing for mathematical treatment of their properties.
Billiard models have been increasingly important in both theoretical and experimental physics, for example as models in statistical mechanics such as the Boltzmann Hypothesis \cite{Szasz}, number theory and the Riemann Hypothesis \cite{BD05}, in room acoustics \cite{Nakano08}, atom optics, where ultracold atoms reflect from laser beams \cite{Montangero09}, optics in dielectric micro-resonators \cite{Nosich08} and in quantum chaos when solving the Helmholtz equation with Dirichlet or Neumann boundary conditions \cite{Stockmann09}. Open billiards are also a useful model for understanding the close correspondence between classical and quantum mechanics \cite{Schomerus05}.

Quantum open billiards were experimentally realized first in flat microwave resonators in the early 90's \cite{Stockmann90,Dietz10} and later in semiconductor nano-structures such as quantum dots \cite{Marcus92,Nakamura04}. Experiments perturbing these systems with small magnetic fields exhibit principal quantum interference effects like weak localization, Altshuler-Aronov-Spivak oscillations and conductance fluctuations, all of which semiclassical theory has arguably succeeded to explain using properties of the underlying classical dynamics \cite{Nakamura04,Jalabert90}. Similarly, in microwave resonators, due to their clean, impurity-free geometry and the tunable coupling strength to the various decay channels, predicted phenomena like resonance trapping have been experimentally observed \cite{Stockmann00}.

\begin{figure}[h]
\begin{center}
\fbox{
\includegraphics[scale=0.22]{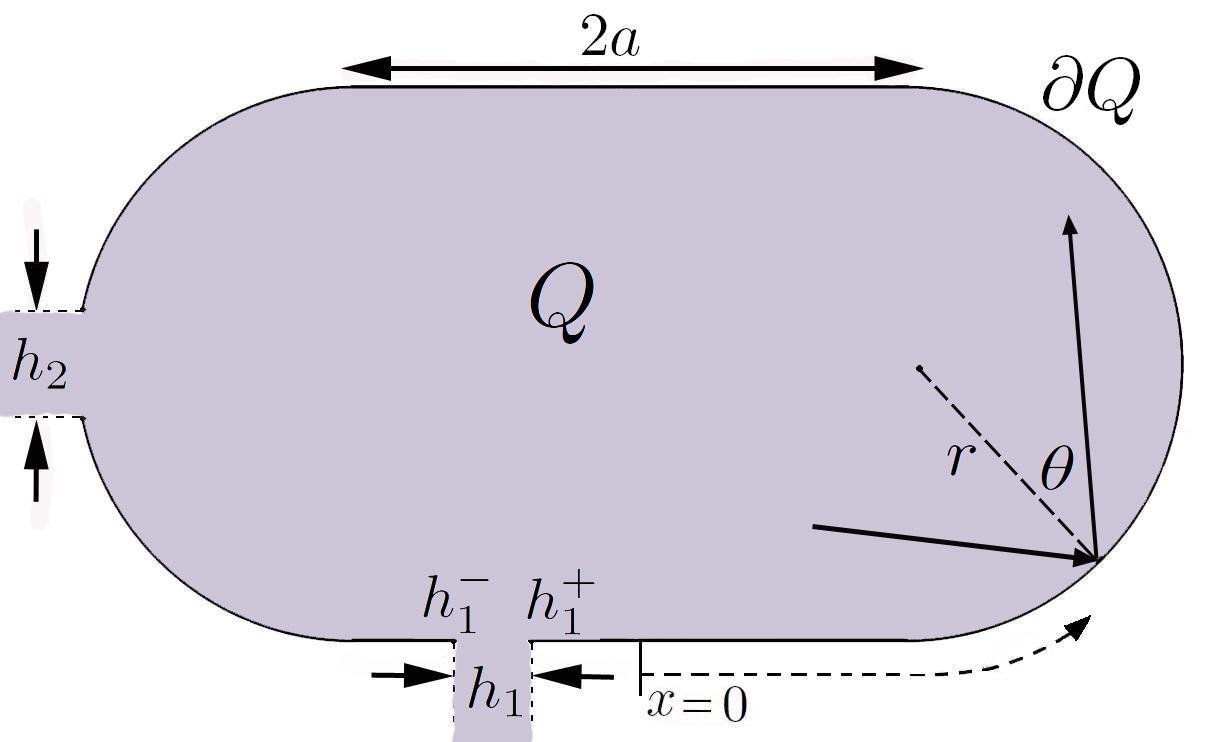}}
\caption{\label{fig:stadium} \footnotesize Stadium billiard with two holes $H_{1}$ and $H_{2}$. The billiard map is parameterized using arc length $0\leq x< 4a+2\pi r$ and velocity parallel to the boundary $v \sin \theta$ with $\theta\in(-\frac{\pi}{2},\frac{\pi}{2})$. The hole's coordinates on the straight segment are defined such that $-a<h_{1}^{-}<h_{1}^{+}<a$.}
\end{center}
\end{figure}

Here we investigate the classical transport of a popular example for the above and other experiments, the stadium billiard with two holes on its boundary (see Figure ~\ref{fig:stadium}). Looking at the phase space of this open system, we find that the predominantly chaotic character of the corresponding closed system is non-trivially affected by the positioning of holes. In particular we find that the transmission and reflection probabilities, when particles are injected from one of the two holes, are qualitatively different at long times depending on the choice of the injecting hole therefore displaying time-dependent asymmetric transport. We give detailed analytical expressions for these distributions and confirm them numerically. Although some work has been done in this direction on the quantum level \cite{Schomerus08}, to the best of our knowledge, there has been no analogous analytic prediction or experimental observation of such an effect. Hence we conclude with a discussion of a possible experimental model with regards to the relevant quantum time scales.

The transport problem is closely related to the escape problem for which we also make new observations. The uniform (Liouville) distribution projected onto the billiard boundary has the form $(2|\partial Q|)^{-1}\mathrm{d}x $ $ \mathrm{d}\sin \theta $ (where $|\partial Q|$ is the perimeter of the billiard while $x$ and $\theta$ are defined in Figure ~\ref{fig:stadium}), and is the most natural choice for an initial distribution of particles. Given such a distribution, the probability $P(t)$ that a particle survives (\textit{i.e.} does not escape through $k$ small holes $H_{i}\in\partial Q$) in a strongly chaotic billiard up to time $t$ decays exponentially $\sim e^{-\gamma t}$ at long times with the exponent to leading order given by \cite{BD07}
\begin{equation}
\gamma= \frac{\sum_{i=1}^{k} h_{i}}{\langle\tau\rangle |\partial Q|},
\end{equation}
where $\langle\tau\rangle=\frac{\pi |Q|}{|\partial Q|v}$ is the mean free path for $2$D billiards, $h_{i}= |H_{i}|$ is the length of each hole, $|Q|$ the area and $v$ the speed of the particles.

The stadium billiard is a chaotic system where the defocusing mechanism guarantees a positive Lyapunov exponent $\lambda$ (exponential separation rate of nearby trajectories) almost everywhere \cite{Bu79}, the exception being a zero-measure family of marginally unstable periodic orbits between the parallel straight segments called Bouncing Ball orbits. They have been shown to lead to an intermittent, quasi-regular behaviour which effectively causes the closed stadium to display some weaker chaotic properties such as an algebraic decay of correlations \cite{Balint08}. Quantum mechanically they cause scarring \cite{Gabriel}, the system is not quantum uniquely ergodic \cite{Hassell10}, an $\hbar$ dependent `island of stability' appears to surround them \cite{GTan} and deviations from random matrix theory (RMT) GOE predictions are observed (especially in the $\Delta_{3}$-statistics) if not treated appropriately (see \cite{Sieber93,Graf92}).

A small hole of size $h_{1}$ on a straight segment, opens the system and the stadium's survival probability $P(t)$ becomes a useful statistical observable. Due to the bouncing ball orbits, $P(t)$ is found to experience a cross-over from the above exponential decay (1) at short times to an algebraic decay $\sim C/t$ at later times \cite{Arm04}. We point out here that the reason for this is that \textbf{the stadium's classical phase space is split into separate regions occupied by `fully-chaotic' and `sticky' orbits, which are responsible for the exponential and algebraic decays respectively}. As an orbit approaches the sticky region in phase space, which surrounds the bouncing ball orbits, it will inevitably escape quickly after it obtains an incidence angle $|\theta| < \arctan \left(\frac{h_{1}}{4 r}\right)$. This is a key point that will be discussed further in the two hole case shortly.

We also remark that due to the splitting of the phase space, there is no justification for an intermediate purely exponential decay, as proposed generically for intermittent systems by Altmann \textit{et al.} (see eq. (25) in \cite{Altman09}), but rather a coexistence of exponential and algebraic decay given by:
\begin{equation}
P(t)=\begin{cases}
\mathrm{irregular}, &   \text{for $t< \hat{t}$}\\
e^{-\gamma t} + \frac{C}{t}, &  \text{for $t>\hat{t}$,}\end{cases}
\end{equation}
\begin{equation}
C= \frac{(3\ln3+4)\Big( (a+h_{1}^{-})^{2} + (a-h_{1}^{+})^{2} \Big)}{4(4a+2\pi r)v},
\end{equation}
with parameters as defined in Figure ~\ref{fig:stadium}, where we have neglected terms of order $t^{-2}$ and $\hat{t}\leq \frac{32 a r}{h_{1}}$ can be found as described in \cite{OreDet09} and guarantees the splitting of the phase space as described above. The `irregular' short-time behaviour is a result of geometry dependent short orbits which become less important if the hole is small. We note that the coefficient of the exponential term in (2) is $1$ since for small holes and times greater than $\approx 1/\lambda$, mixing causes the system to forget its initial state and therefore the probability decays as a Poisson process.
\begin{figure}[h]
\begin{center}
\fbox{
\includegraphics[scale=0.33]{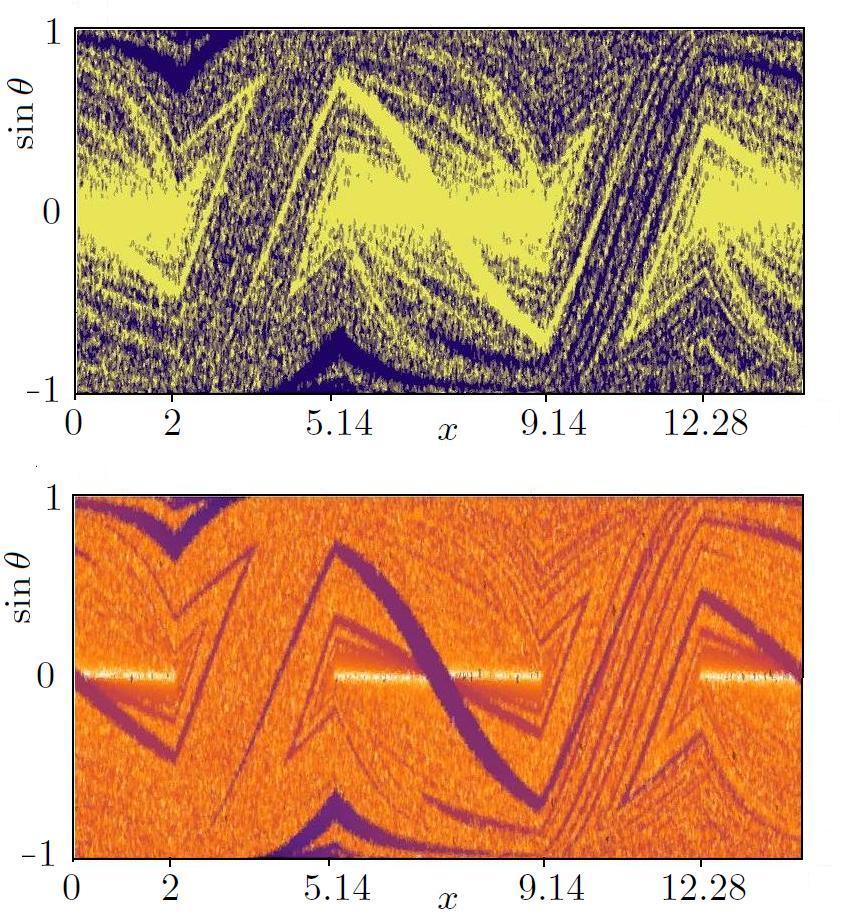}}
\caption{\label{fig:colour} \footnotesize (Colour online) Phase space of open stadium with $2$ holes. \emph{Top}: Initial conditions which will escape through hole $H_{1}$ are shown in light yellow while those escaping through hole $H{2}$ in dark blue. \emph{Bottom}: Colour grading of initial conditions going from purple, to orange, to white corresponding to short, medium and long escape times. ($a=2$, $r=1$, $h_{i}=0.5$, $h_{1}^{+}=0.25$).}
\end{center}
\end{figure}

\begin{figure}[h]
\begin{center}
\fbox{
\includegraphics[scale=0.21]{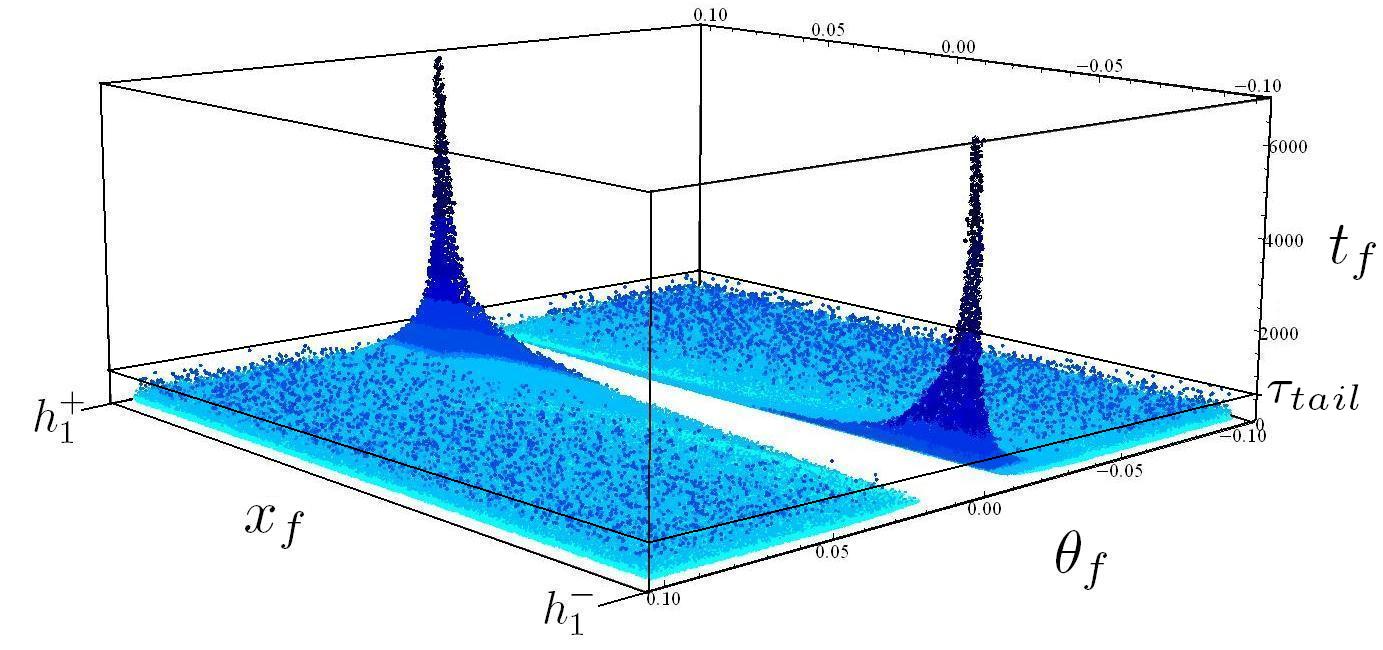}}
\caption{\label{fig:escape} \footnotesize (Colour online) 3D plot of the final (escape) coordinates and time of escape $(x_{f},\theta_{f},t_{f})$ for the case of entry and exit through $H_{1}$. Only in this case are the $2$ dark spikes observed. For other entry/exit combinations, a uniform `carpet' with an exponentially decaying number of particles for larger $t$ would be observed. The colour grading emphasizes the magnitude of the survival time of each orbit. ($a=2$, $r=1$, $h_{i}=0.2$, $h_{1}^{+}=0.1$). $\tau_{tail}\approx631.85$ is explained in Figure ~\ref{fig:numerics}.}
\end{center}
\end{figure}

Consider now the case of the stadium with two holes as shown in Figure ~\ref{fig:stadium}. In Figure ~\ref{fig:colour} we plot in the top panel a picture of the phase space, showing in different colours, the different sets of initial conditions which eventually exit through each hole. The bottom panel shows the time scales of escape as noted in the caption. We notice that the phase space is again separated, as described above, and that the sticky, long-surviving orbits escape only through the hole on the straight segment $H_{1}$. Restricting the initial density of particles to one of the holes defines the transport problem and establishes the schematic setup of quantum dots and microwave cavities, where particles/waves are injected through one of the holes and allowed to escape through either, thus creating a direct link with experiment. Looking at the spatial distribution of the final (escape) coordinates $(x_{f},\theta_{f})$ (see Figure ~\ref{fig:escape}) we observe that the long surviving orbits entering and subsequently exiting through $H_{1}$, unlike in the other possible entry/exit combinations, accumulate on the edges of the hole $x_{f}=h_{1}^{\pm}\mp\delta$, $(\delta\ll1)$ and have small angles $\theta_{f}$. Note that $(x_{f},\theta_{f})\rightarrow (h_{1}^{\pm},0^{\pm})$ as the time of escape $t_{f}\rightarrow \infty$. This further confirms the splitting of the phase space, but also that the classical \emph{spatial} distribution of exiting particles has a well defined time-dependent character, which only exists in the situation described and plotted in Figure ~\ref{fig:escape}.

We define transmission and reflection survival probabilities by $P_{i}^{j}(t)$ and $P_{i}^{i}(t)$ respectively ($i, j =1,2$), such that
\begin{equation}
P_{i}^{j}(t)=P(x_{1}\ldots x_{\mathcal{N}}\notin H \big| x_{0}\in H_{i}, x_{f} \in H_{j}),
\end{equation}
where $H=H_{1}\cup H_{2}$, $\mathcal{N}(x_{0},t)$ is the number of collisions with the boundary up to time $t$ and $x_{n}$ denotes the position of the particle at the $n$th collision. Hence, $P_{1}^{2}(t)$ is the probability that a particle injected from hole $H_{1}$ will survive until time $t$ given that it will escape through hole $H_{2}$. It follows that only $P_{1}^{1}(t)$, out of the four possible survival distributions, has an algebraic tail, while the other three decay exponentially with an escape rate given by $\gamma= \frac{h_{1}+h_{2}}{\langle\tau\rangle |\partial Q|}$.
Also, for $P_{1}^{1}(t)$, at $t\gg1$ and $\theta\ll1$, only particles starting near the edges of the hole are not immediately reflected back into the hole. Therefore, the extra constraint $|\theta|>\arctan \left|\frac{h_{1}^{\pm}-x_{0}}{4 r}\right|$ ($\pm$ depending on the sign of $\theta$) gives $P_{1}^{1}(t)$ an algebraic tail $\mathcal{O}(t^{-2})$, as expected in integrable scattering problems.
In summary the total survival probability $P_{i}(t)$, where the subscript $i$ indicates the injecting hole, is given by:
\begin{align}
P_{1}(t)&=e^{-\gamma t} + \frac{D}{t^{2}}=\wp_{1}^{1}\overbrace{ \left(e^{-\gamma t}+ \frac{D}{\wp_{1}^{1}t^{2}}\right)}^{P_{1}^{1}(t)} +\wp_{1}^{2}\overbrace{e^{-\gamma t}}^{P_{1}^{2}(t)},\\
P_{2}(t)&=e^{-\gamma t} =\wp_{2}^{2}\overbrace{e^{-\gamma t}}^{P_{2}^{2}(t)} +\wp_{2}^{1}\overbrace{e^{-\gamma t}}^{P_{2}^{1}(t)},
\end{align}
for $t>\hat{t}$, where the $\wp_{i}^{j}$ are the respective asymptotic ($t\rightarrow\infty$) reflection and transmission coefficients. Notice that $\wp_{i}^{1}+\wp_{i}^{2}=1$ due to flux conservation, and $\wp_{1}^{2}=\wp_{2}^{1}$ due to time-reversal symmetry. $D$ is given by a similar calculation to \cite{OreDet09}:
\begin{equation}
D=\frac{ r(3\ln3+4)\Big( (a+h_{1}^{-})^{2} + (a-h_{1}^{+})^{2} \Big)}{2h_{1}v^{2}}.
\end{equation}

\begin{figure}[h]
\begin{center}
\fbox{
\includegraphics[scale=0.23]{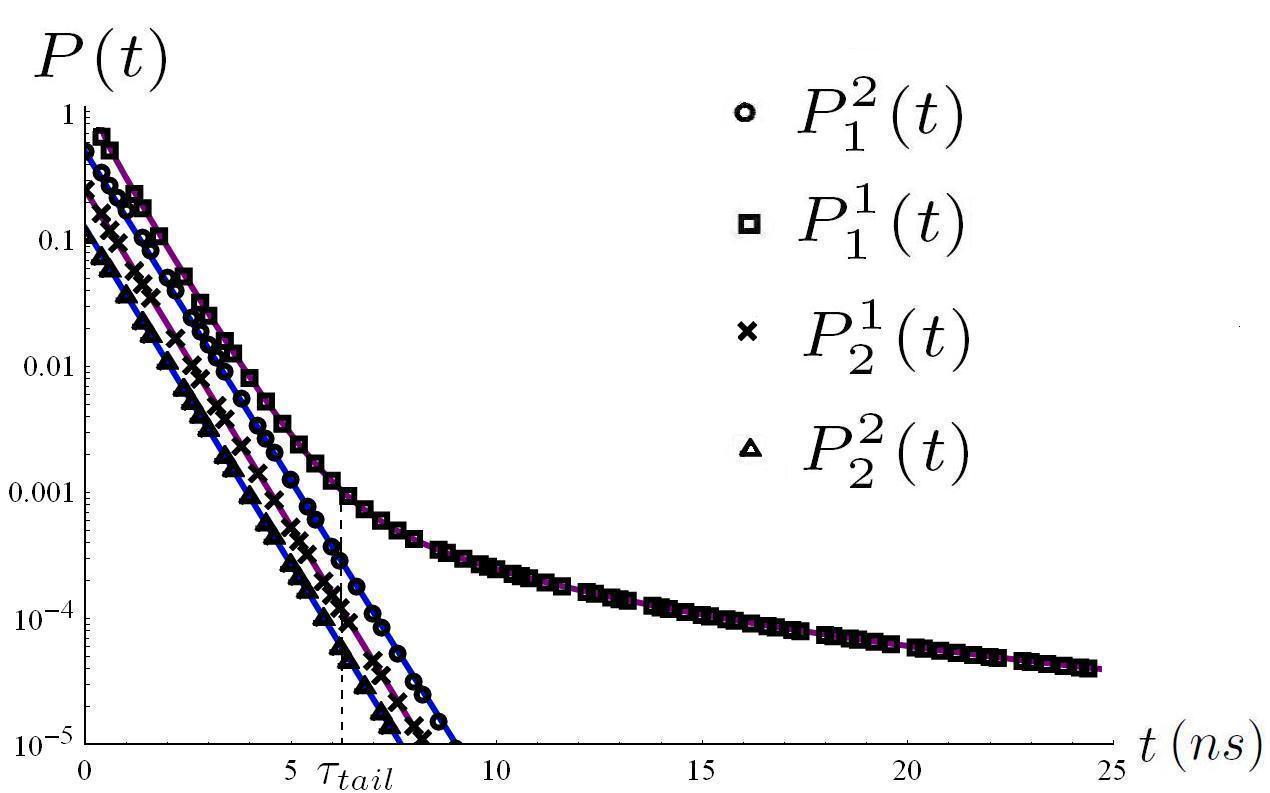}}
\caption{\label{fig:numerics} \footnotesize Slightly offset plots comparing numerical simulations of the conditional survival probabilities (empty circles and squares) with the analytic expressions (5) and (6) as functions of time $t$ in $ns$. The simulations consist of $10^{9}$ particles with stadium parameters given by: $a=2$ $\mu m$, $r=1$ $\mu m$, $h_{i}=0.2$ $\mu m$ and $h_{1}^{-}=0$. $\tau_{tail}\approx 6.315$ $ns$ is the large solution of $e^{-\gamma t} = \frac{D}{\wp_{1}^{1}t^{2}}$, where $\wp_{1}^{1}\approx 0.5594$ was calculated numerically.}
\end{center}
\end{figure}

In Figure ~\ref{fig:numerics}, we plot the four conditional distributions $P_{i}^{j}(t)$ as functions of time $t$, and find an excellent agreement with the analytical results summarised in equations (5-6). We emphasize that the power law decay of $P_{1}^{1}(t)$ is due to the geometric asymmetry of the hole's positions which exploit the marginally unstable bouncing ball orbits as to force a preference of escape through $H_{1}$.
Furthermore, the splitting of the phase space into fully chaotic and sticky regions renders the later \textbf{inaccessible} to particles injected through $H_{2}$.
This would not have been be the case if both $H_{1}$ and $H_{2}$ were placed on a straight or curved segment. \textbf{This important observation offers the simplest possible example where a classically fully chaotic system exhibits time-asymmetric transport when opened.}
This is expected to be relevant to many other intermittent systems. Also, the variety of options with regards to hole positions and sizes and system parameters offers ways of calibrating and controlling these classical distributions and hence encourages the possibility of experimental observation of the quantum analogue which we discuss next.

At low temperatures ($\sim 15$ $mK$), electronic transport through the gate electrodes (openings) of a 2D electron gas (quantum dot) is ballistic \cite{Marcus92,Nakamura04}.
For typical semiconductor nano-structure parameters, the time scale $\tau_{tail}$ at which the above observed algebraic tail becomes visible (see Figures ~\ref{fig:escape} and ~\ref{fig:numerics}) is of the order of a nanosecond (assuming an electron speed $v\approx 10^{5}$ $m s^{-1}$). This is slightly larger than the predicted Ehrenfest time $\tau_{E}= \lambda^{-1}\ln \left(1/\hbar\right)$ for chaotic systems \cite{Schomerus05} (the time scale at which quantum interference effects become apparent $\approx 0.3$ $ns$), and thus at first instance suggests that direct observation of a quantum difference in transmission and reflection survival probabilities is unlikely in existing devices. We find that by varying the size and hole positions of the dot (while remaining in the ballistic regime) it is possible to calibrate and reduce $\tau_{tail}$ by a whole order of magnitude. A good way to do this is by elongating the stadium slightly such that $a/r\approx5$ and by placing $H_{1}$ at the very edge of the straight segment. However, since the nature of chaos lies in orbital instability, \textbf{the Ehrenfest time $\tau_{E}$ calculated from the average Lyapunov exponent $\lambda$ does not faithfully represent quantum spreading of the near-bouncing ball orbits}, which are responsible for the algebraic decay of $P_{1}^{1}(t)$ noted here. For these, the finite-time local Lyapunov exponent is zero \cite{Casati05}, therefore leading to a much longer validity and persistence of the classical description of the sticky region in phase space. In fact, this region could be thought of as an $h_{1}$ depended fictitious island of stability in which loss of quantum-to-classical correspondence is much slower, resembling that in mixed systems, such that $\tau_{E}\propto\hbar^{-1/\beta}$ \cite{Fishman87}. Therefore, experimental realization of the asymmetric transport (5) and (6) would not only \textbf{confirm classical-quantum correspondence of the time-scales proposed in this particular system, but also emphasize the need for quantum corrections for transport in intermittent chaotic systems}.

Suppose we apply a time-dependent voltage $V(t)$ across the gates of the stadium heterostructure such that the incoming current $I_{i}^{in}(t)$ through hole $H_{i}$ is proportional to $V(t)$. Then the charge exiting through each hole will follow the driving current with a lag-time $\tau$ which is distributed according to (5) or (6) appropriately. This can be modeled by
\begin{equation}
I_{j}(t)= (-1)^{i+j}\wp_{i}^{j}\int_{0}^{\infty}I_{i}^{in}(t-\tau)\frac{\textrm{d} P_{i}^{j}(\tau)}{\textrm{d} \tau} \mathrm{ d} \tau,
\end{equation}
where $i$ and $j$ indicate the injecting and exiting hole respectively. The observed, net current through the system is thus given by $I_{i}^{net}(t)=I_{i}^{in}(t)+I_{1}(t)+I_{2}(t)$. Because the probability density $\frac{\textrm{d} P_{1}^{1}(\tau)}{\textrm{d} \tau}$ is slightly skewed to the right, relative to the other densities, the two observables $I_{1}^{net}(t)$ and $I_{2}^{net}(t)$ will differ by
\begin{align}
\wp_{1}^{1}\int_{0}^{\infty}  \frac{\textrm{d}I^{in}(t-\tau)}{\textrm{d} \tau}\left(P_{1}^{1}(t)-P_{2}^{2}(t)\right)\textrm{d} \tau.
\end{align}
For experimental observation we propose using a square wave signal $V(t)=V_{0}\left(1+ sign(\sin\omega t)\right)$ such that $\omega> \pi/\tau_{tail}$ as to accentuate the power-law contribution of $P_{1}^{1}(t)$.
Quantum interference effects such as universal conductance fluctuations may be statistically removed since the skewness of $\frac{\textrm{d} P_{1}^{1}(\tau)}{\textrm{d} \tau}$ is to leading order geometry dependent through the constant $D$ in (7). In experiments of course, one should make sure that the excess density of charged particles within the dot is always low enough as to avoid a build up of an internal electric field which would effectively destroy the fictitious island of stability (sticky region) enclosing the bouncing ball orbits. For microwave billiards this is not an issue.

To conclude, we have investigated the classical dynamics of the chaotic stadium billiard with one and two holes, and have found that the transmission and reflection survival distributions of the latter case can have algebraic and exponential decays observed in the same classically ergodic geometry. We have identified the reason for this being the hole's asymmetric positioning on the straight segment of the billiard, which essentially splits the classical phase space of the system, rendering the sticky region surrounding the bouncing ball orbits inaccessible to chaotic orbits. As a result, the transmission and reflection survival distributions are qualitatively different. We have obtained analytic expressions, confirmed them numerically and propose that observation of this classical result in semiconductor nano-structures (quantum dots) or microwave cavities would improve our understanding of classical to quantum correspondence in transport problems.
Specifically it would imply that the Ehrenfest time and more generically quantum chaos predictions have correction terms subject to the underlying classical dynamics of the corresponding \textbf{open} systems. Furthermore, this would possibly introduce new ways of calibrating and controlling transport through complex systems by utilizing the sticky (non-mixing) channels of intermittent systems.

We would like to thank M. Sieber, R. Schubert, M. Fromhold, A. Micolich and H. Schomerus for helpful discussions and OG's EPSRC Doctoral Training Account funding.

\end{document}